\documentclass[
aip,
reprint,
superscriptaddress,
amsmath,
amssymb]{revtex4-1}

\usepackage{graphicx}
\usepackage{dcolumn}
\usepackage{bm}

\usepackage[utf8]{inputenc} 
\usepackage[T1]{fontenc}    
\usepackage{hyperref}       
\usepackage{url}            
\usepackage{booktabs}       
\usepackage{amsfonts}       
\usepackage{nicefrac}       
\usepackage{microtype}      
\usepackage{natbib}
\usepackage{makecell}
\usepackage{comment}
\usepackage{xcolor}
\usepackage{mathptmx}
\usepackage{etoolbox}
\usepackage{upgreek}

\makeatletter
\def\@email#1#2{%
 \endgroup
 \patchcmd{\titleblock@produce}
  {\frontmatter@RRAPformat}
  {\frontmatter@RRAPformat{\produce@RRAP{*#1\href{mailto:#2}{#2}}}\frontmatter@RRAPformat}
  {}{}
}%
\makeatother
\begin{document}

\title{Nonlinear near-field spectroscopy of exciton-polaritons in a van der Waals layered waveguide}

\author{Valeriy I. Kondratyev}
\affiliation{School of Physics and Engineering, ITMO University, Saint Petersburg 197101, Russia}

\author{Vanik Shahnazaryan}
\affiliation{School of Physics and Engineering, ITMO University, Saint Petersburg 197101, Russia}

\author{Mikhail Tyugaev}
\affiliation{School of Physics and Engineering, ITMO University, Saint Petersburg 197101, Russia}

\author{Tatyana V. Ivanova}
\affiliation{School of Physics and Engineering, ITMO University, Saint Petersburg 197101, Russia}

\author{Ivan E. Kalantaevskii}
\affiliation{School of Physics and Engineering, ITMO University, Saint Petersburg 197101, Russia}

\author{Dmitry V. Permyakov}
\affiliation{School of Physics and Engineering, ITMO University, Saint Petersburg 197101, Russia}

\author{Ivan V. Iorsh}
\affiliation{Department of Physics, Engineering Physics and Astronomy, Queen’s University, Kingston, Ontario K7L 3N6, Canada}

\author{Anton K. Samusev}
\affiliation{Experimentelle Physik 2, Technische Universität Dortmund, 44227 Dortmund, Germany}

\author{Vasily Kravtsov}
\email{vasily.kravtsov@metalab.ifmo.ru}
\affiliation{School of Physics and Engineering, ITMO University, Saint Petersburg 197101, Russia}


\begin{abstract}
\noindent
Layered van der Waals materials offer novel opportunities for on-chip waveguiding and development of integrated photonic circuits.
In the strong light--matter coupling regime, their nonlinear response can be significantly enhanced, which is crucial for developing active photonic devices.
However, probing the nonlinearity of waveguide modes in subwavelength-thick structures is challenging as they are not directly accessible from far-field.
Here we apply a novel nonlinear near-field spectroscopic technique based on a GaP solid immersion lens and femtosecond laser excitation to study nonlinearity of guided modes in monolayer WS$_2$ encapsulated in hBN under the strong light--matter coupling regime.
We reveal formation of exciton-polaritons with $\sim 50$~meV Rabi splitting and demonstrate a pump-induced transition from strong to weak coupling.
Our results show that exciton resonance saturation and broadening lead to an efficient nonlinear response of guided polaritons, which can be employed for developing compact van der Waals photonic switches and modulators.
\end{abstract}


\maketitle

\noindent
Two-dimensional transition metal dichalcogenides (TMDs) provide a promising material platform for future compact and chip-compatible photonic and optoelectronic elements~\cite{Mak2016}.
In their monolayer limit, TMDs are direct bandgap semiconductors with large (100s of meV) exciton binding energies~\cite{Hanbicki2015}, relatively high exciton oscillator strength~\cite{Li2014}, and efficient exciton--exciton interaction~\cite{Shahnazaryan2017}.
This leads to strong photoluminescence (PL), room-temperature operation, and significant nonlinear response of excitons in TMD monolayers providing possibilities for all-optical control~\cite{Sun2016, Chen2020}.

As van der Waals materials, TMDs can be easily encapsulated into high-quality multi-layer crystals of hexagonal boron nitride (hBN), which proved to be a powerful approach towards controlling and improving their excitonic response.
Creating a vertical hBN/TMD/hBN heterostack can, for example, significantly narrow the exciton linewidth down to its radiative limit~\cite{Ajayi2017, Cadiz2017, Martin2020}, increase the on-resonance reflectivity~\cite{Back2018, Scuri2018}, enhance the PL response~\cite{Hoshi2017}, and extend the exciton diffusion length by an order of magnitude~\cite{Zipfel2020}.
Recently, there has been increasing interest in using thin hBN/TMD/hBN structures for waveguiding, where PL coupling to guided modes can be optimized for certain hBN thickness~\cite{Frank2022, Khelifa2023, LaGasse2023, Glazov2024}, while nanostructuring allows fabrication of compact metasurfaces, resonators, and grating couplers~\cite{Gelly2023} creating a new material platform for integrated photonic devices.

For further development of active integrated photonic elements based on hBN/TMD/hBN, it is crucial to engineer an enhanced and controllable nonlinear response.
A promising way to achieve it is via exciton-polariton formation in the strong light--matter coupling regime~\cite{Khitrova1999, Schneider2018}.
However, most implementations of nonlinear strongly-coupled systems based on TMD monolayers to date have relied on the use of additional photonic structures such as distributed Bragg reflectors~\cite{Barachati2018, Tan2020, Zhao2024}, nanostructured metasurfaces~\cite{Kravtsov2020, khestanova2024electrostatic, Sortino2025}, or slabs of high-index dielectrics~\cite{Benimetskiy2023}, which requires additional and often expensive fabrication processes.
Unpatterned hBN/TMD/hBN stacks can support guided exciton-polaritons, but it is challenging to access these states in experiment due to their high wavevectors inaccessible for traditional far-field spectroscopy techniques. 
To date, exciton-polaritons in these structures have been probed only in the linear optical regime and with multilayer TMDs~\cite{Lee2024}. 

In this work, we demonstrate that formation of nonlinear exciton-polaritons can be achieved at room conditions in a simple all-van-der-Waals hBN/WS$_2$/hBN heterostructure with optimized hBN thickness.
In order to experimentally access the resulting guided exciton-polariton modes and probe their nonlinear behavior, we use a novel spectroscopic technique based on evanescent wave coupling through a high-index solid immersion lens (SIL) in combination with femtosecond laser pulse excitation.
The observed guided exciton-polaritons exhibit large Rabi splitting of $\sim 50$~meV, which is controllably reduced in experiment via power-dependent resonant optical pumping, leading to a strong-to-weak coupling transition at pump fluence of $\gtrsim 10~\upmu \mathrm{J/cm^2}$.
We support our findings with a theoretical model of polariton--polariton interaction that reveals the key contribution of pump-induced Rabi splitting collapse and exciton broadening to the observed nonlinearity.
Our results suggest that hBN/TMD/hBN heterostacks are a promising room-temperature platform for future studies of nonlinear guided polaritons and associated effects such as frequency conversion, self-phase modulation, and soliton formation.

\begin{figure*}[t]
    \centering
    \includegraphics[width=\textwidth]{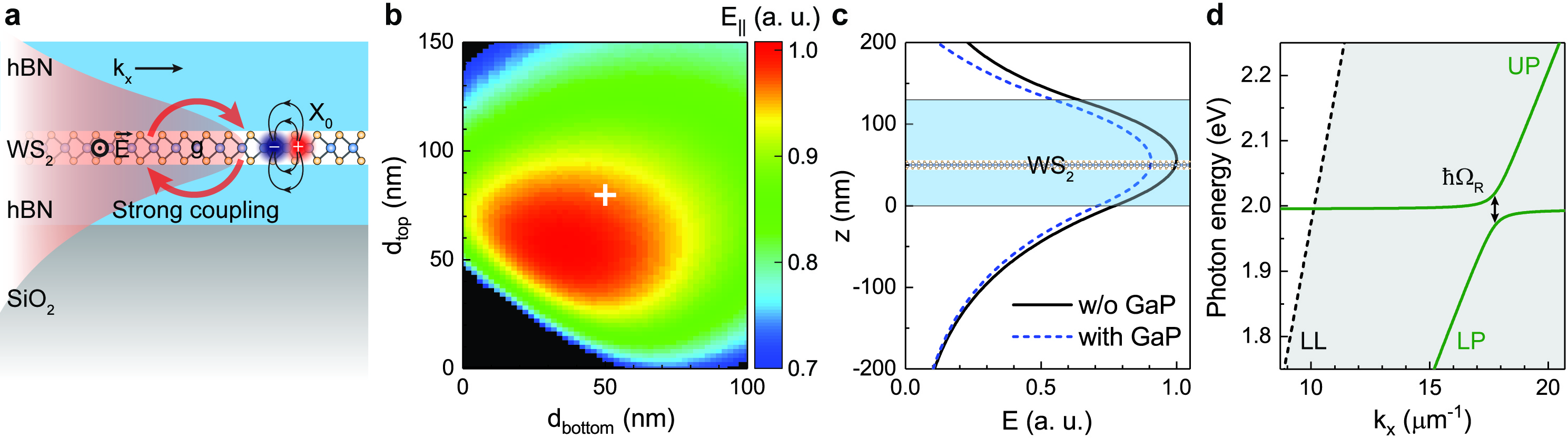}
    \caption{
    \textbf{Strong light--matter coupling in an all-van-der-Waals hybrid waveguide.}
    (a) Schematic illustration of a layered waveguide consisting of monolayer WS$_2$ encapsulated in two multilayer hBN slabs placed on a SiO$_2$ substrate.
    (b) In-plane field amplitude evaluated at the monolayer WS$_2$ position in the layered structure as a function of the bottom and top hBN layer thickness.
    (c) Distribution of the E-field amplitude along the out-of plane z-axis for the lowest-order TE mode in a hBN/WS$_2$/hBN/SiO$_2$ structure (black curve) and its modification in the presence of GaP solid immersion lens placed 200 nm above the structure.
    (d) Calculated dispersion of the lowest-order TE waveguide mode (green curves) showing Rabi splitting and formation of lower (LP) and upper (UP) polariton branches, together with light line (LL, black dashed line).
    }
    \label{fig:Idea}
\end{figure*}

Fig.~\ref{fig:Idea}a shows a schematic of the structure considered in our work, which consists of a monolayer WS$_2$ sandwiched between two multi-layers of hBN placed on a SiO$_2$ substrate.
Because of the reasonably high in-plane refractive index of hBN at visible frequencies of $\sim 2.1$, the structure can be viewed as a high-index dielectric slab waveguide that supports propagation of non-radiating optical modes.
These modes couple to the exciton resonance in monolayer WS$_2$ with a coupling strength $g \propto f/\sqrt{V}$, where $f$ is the exciton oscillator strength, and $V$ is the optical mode volume.
For sufficiently high $g$, the conditions of strong light-matter coupling can be satisfied, and the structure will support guided exciton-polariton modes.

In order to find the optimal geometry of the hBN/WS$_2$/hBN structure for achieving the strong coupling regime, we consider the lowest-order waveguide TE mode and calculate~\cite{OMS} its E-field amplitude at the monolayer location for different thicknesses of the bottom and top hBN layers, with the results shown in Fig.~\ref{fig:Idea}b.
As seen in the plot, the dependence of the E-field amplitude is non-monotonic and peaks at certain thickness values.
For the experiment, we choose 50~nm and 80~nm for the bottom and top hBN thickness, respectively, corresponding to a near-optimal E-field amplitude that minimizes the mode volume $V$ and thus maximizes the coupling strength $g$.
The E-field profile along the out-of-plane direction (z-axis) for the chosen parameters is shown in Fig.~\ref{fig:Idea}c (black solid curve) and has an asymmetric shape due to the presence of SiO$_2$ beneath the waveguide.

The chosen geometry is expected to achieve strong coupling between  the TE waveguide mode and A-exciton resonance in monolayer WS$_2$ at E$_\mathrm{X} = 1.995$~eV.
The resulting calculated dispersion is shown in Fig.~\ref{fig:Idea}d, where the waveguide mode (green curve) lies significantly below the light line (LL, black dashed curve) and is split into the lower (LP) and upper (UP) polariton branches with Rabi splitting $\hbar\Omega_\mathrm{R} \sim 50$~meV.
We note that, in contrast to TM surface polaritonic states considered previously~\cite{Epstein2020}, here we study TE modes corresponding to bulk guided polaritons.
Direct experimental observation of guided polaritons is challenging as their dispersion lies below the light line (gray-colored area in Fig.~\ref{fig:Idea}d), which forbids direct coupling to far-field light.

\begin{figure*}[t]
    \centering
    \includegraphics[width=\textwidth]{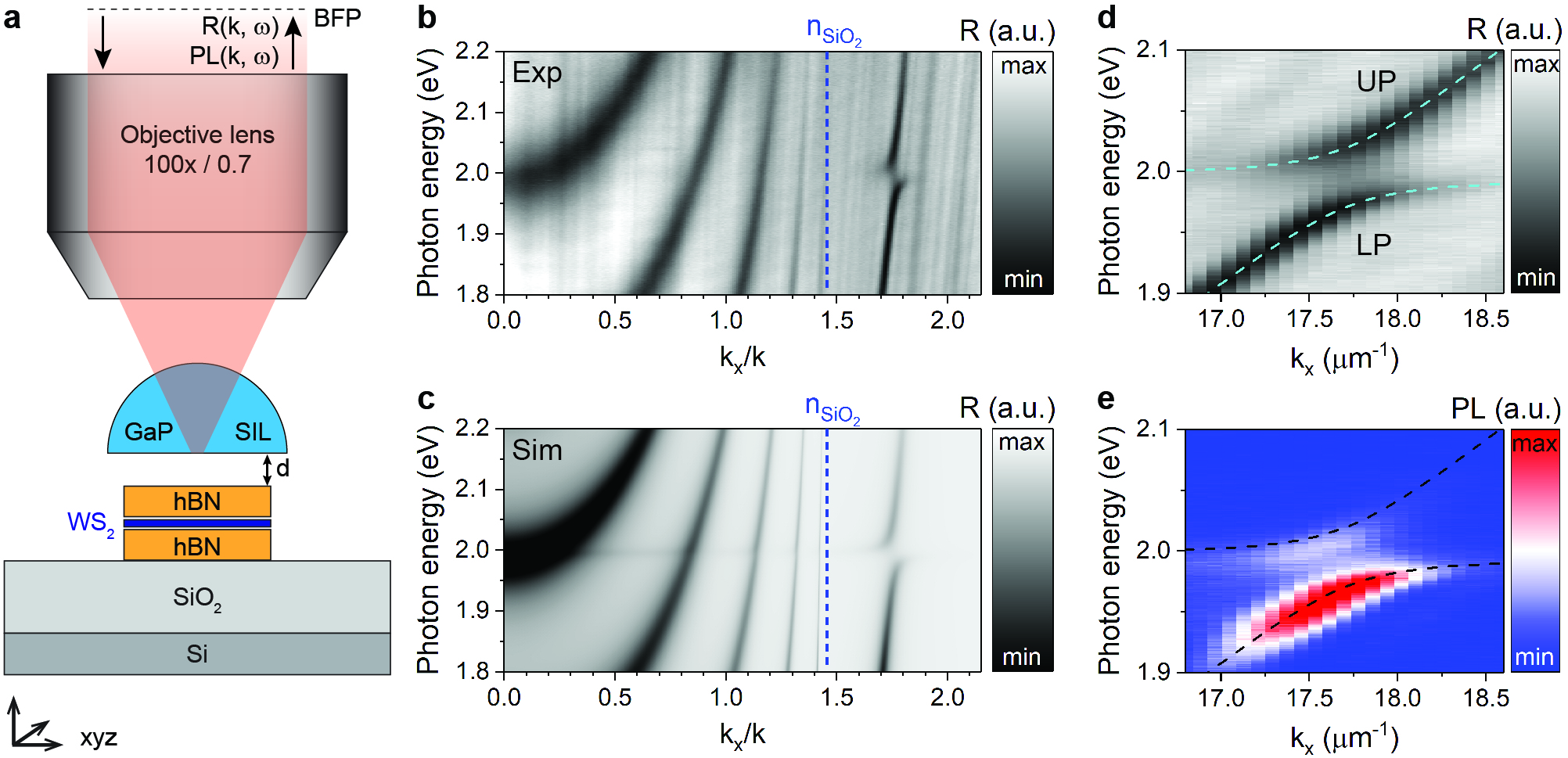}
    \caption{
    \textbf{Measurement of guided exciton-polaritons in an all-van-der-Waals hybrid waveguide.}
    (a) Schematic illustration of the experimental configuration for angle-resolved measurements below the light line including a GaP solid immersion lens with tunable lens--sample gap and high-NA microscope objective.
    (b) Measured optical reflection as a function of photon energy and relative in-plane wavevector component k$_x$/k. The region corresponding to guided modes lies to the right of the blue dashed line indicating the refractive index of SiO$_2$.
    (c) Simulated optical reflection as a function of photon energy and relative in-plane wavevector component k$_x$/k.
    (d) Measured wavevector-resolved reflection spectra around the exciton resonance showing anti-crossing behavior and formation of lower (LP) and upper (UP) polariton branches.
    (e) Corresponding wavevector-resolved PL spectra.
    }
    \label{fig:Linear}
\end{figure*}

In order to access guided polaritons in the experiment, we fabricate hBN/WS$_2$/hBN samples (see Supplementary Note 1) and employ a home-built setup based on a high-index SIL~\cite{Permyakov2018, Permyakov2021, kondratyev2023probing} as described in Supplementary Note 2 and schematically illustrated in Fig.~\ref{fig:Linear}a.
The SIL is brought into the near field of the sample, with SIL--sample gap of less than 200~nm controlled by precise piezo positioners.
The use of a GaP SIL with refractive index of $n_\mathrm{GaP} = 3.3$ combined with a high-aperture (NA = 0.7) microscope objective allows us to extend the overall numerical aperture of the setup to $\mathrm{NA_{eff}} = \mathrm{NA} \cdot n_\mathrm{GaP} = 2.3$.
At the same time, the presence of SIL affects the E-field distribution inside the sample only slightly as shown in Fig.~\ref{fig:Idea}c (dashed blue curve).
To obtain the dispersion of leaky and guided modes supported by the hBN/WS$_2$/hBN hybrid waveguide, we project the back focal plane (BFP) of the objective onto a slit of a spectrometer and measure the resulting angle-resolved reflection (R) and photoluminescence (PL) spectra.

Fig.~\ref{fig:Linear}b shows typical data measured on our hBN/WS$_2$/hBN sample, with $x$ and $y$ axes corresponding to the relative in-plane wavevector component $k_x/k$ and photon energy $\hbar\omega$, respectively, and the color scale corresponding to the reflected signal intensity.
For small wavevectors $k_x/k < n_\mathrm{SiO_2}$ (indicated with dashed blue line), we observe leaky modes as fringes in the reflection spectra corresponding to Fabry--Perot resonances in the layered substrate.
In the region of high wavevectors below the light line $k_x/k > n_\mathrm{SiO_2}$, we observe a TE waveguide mode, which exhibits visible splitting at the WS$_2$ A-exciton resonance energy of 1.995~eV.
The observed features are in a very good agreement with the results of transfer matrix calculations plotted in Fig.~\ref{fig:Linear}c, which account for radiative $\gamma_\mathrm{r} = 1.8$~meV and nonradiative $\gamma_\mathrm{nr} = 10.0$~meV exciton relaxation rates.
In calculations (see Supplementary Note 3), we also account for scattering losses in the waveguide via an additional imaginary part of the hBN refractive index $\delta n = 0.025i$, which yields a relaxation rate $\gamma_\mathrm{ph} = 19.4$~meV near the WS$_2$ exciton resonance.

In Fig.~\ref{fig:Linear}d, we provide a more detailed plot for experimental reflection data near the WS$_2$ exciton resonance, with x-axis corresponding to the absolute value of in-plane wavevector component $k_x$.
The waveguide mode and exciton resonance exhibit a clear anticrossing, forming a lower (LP) and upper (UP) polariton branches.
From fits of the data with the coupled oscillators model~\cite{Skolnick1998} (cyan dashed curves), we extract the exciton--photon coupling strength $g = 24.5$~meV.
This value satisfies the two conditions for strong exciton--photon coupling: $g > \sqrt{(\gamma_\mathrm{ph}^2 + \gamma_\mathrm{exc}^2)/2} \simeq 16.1$~meV, $g > |\gamma_\mathrm{ph} - \gamma_\mathrm{exc}|/2 \simeq 3.8$~meV, where $\gamma_\mathrm{exc} = \gamma_\mathrm{r} + \gamma_\mathrm{nr} = 11.8$~meV is the total exciton relaxation rate.

We note that the exciton--photon coupling strength in our system can be sensitively controlled via varying the SIL--sample distance~\cite{kondratyev2023probing}.
On the one hand, bringing the SIL closer to the sample surface increases radiative losses and distorts the optical field distribution, leading to a decrease in coupling strength.
On the other hand, for larger SIL--sample distances, sensitivity to evanescent waves in the sample near field is reduced, and spectral features corresponding to guided modes below the light line become indiscernible.
Therefore, in the experiment we optimize the SIL--sample distance to maximize both the exciton--photon coupling strength and reflectivity contrast for guided modes.
From comparison of measured angle-resolved reflection data with simulation results (see Supplementary Note 3), we estimate the SIL--sample distance $d$ in our experimental configuration to be in the $150-160$~nm range.
The formation of distinct LP and UP exciton-polariton branches in the waveguide dispersion is also observed in angle-resolved PL spectra shown in Fig.~\ref{fig:Linear}e and Supplementary Note 4, confirming that the system is in the strong coupling regime.

\begin{figure*}[t]
    \centering
    \includegraphics[width=\textwidth]{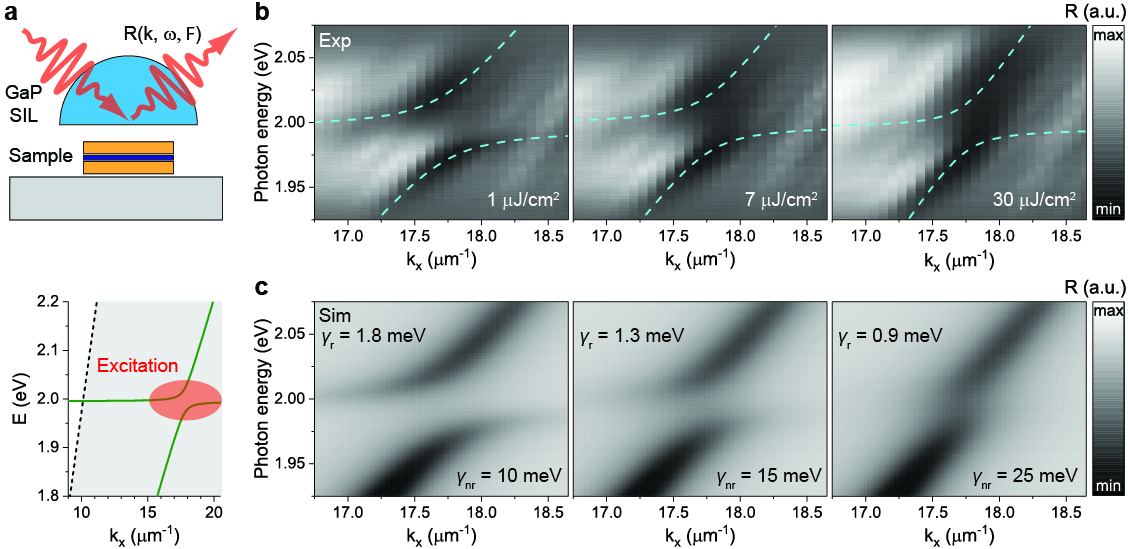}
    \caption{
    \textbf{Measurement of the nonlinear response of guided exciton-polaritons.}
    (a) Schematic of the measurement, with femtosecond laser pulses exciting exciton-polaritons in the sample through SIL (top) resonantly in both photon energy and wavevector (bottom).
    (b) Experimental exciton-polariton dispersions measured as wavevector-resolved optical reflection spectra at 3 selected values of pump fluence: 1 $\upmu$J/cm$^2$ (left panel), 7 $\upmu$J/cm$^2$ (middle panel), and 30 $\upmu$J/cm$^2$ (right panel). 
    (c) Simulated exciton-polariton dispersions for 3 different values of exciton radiative decay rate: 1.8~meV (left panel), 1.3~meV (middle panel), and 0.9~meV (right panel).
    }
    \label{fig:Nonlinear}
\end{figure*}

The observed polaritons are expected to exhibit a significantly nonlinear behavior inherited from the efficient exciton--exciton interactions in monolayer TMD.
Here we probe the nonlinear optical response of guided exciton-polaritons in our hBN/WS$_2$/hBN waveguide directly by combining SIL-based near-field spectroscopy with femtosecond laser pulse excitation.
As described in Supplementary Note 2 and schematically illustrated in Fig.~\ref{fig:Nonlinear}a (top panel), we couple femtosecond laser pulses into the sample through GaP SIL resonantly both in photon energy and wavevector (bottom panel, red ellipse) and measure angle-resolved reflectivity spectra at variable pump fluence.
To accurately estimate pump fluence at the monolayer location, we account for the local field at the estimated experimental SIL--sample gap of $d = 154$~nm using transfer matrix simulations.

Fig.~\ref{fig:Nonlinear}b shows typical measured data (color scale) for 3 selected pump fluence values together with fits of polariton branches via the model of 2 coupled oscillators (dashed curves).
As seen in the plots, at a low pump fluence value of $\sim 1~\upmu \mathrm{J/cm^2}$ (left panel), the polariton dispersion near the exciton resonance is very similar to that for the linear case presented in Fig.~\ref{fig:Linear}d.
From fitting the low-fluence reflection data with coupled oscillators model we extract the exciton--photon coupling strength $g = 25.5$~meV, which corresponds to Rabi splitting of $\sim 50$~meV.
At a moderate pump fluence value of $\sim 7~\upmu \mathrm{J/cm^2}$ (middle panel), the experimentally observed polariton dispersion branches slightly broaden and exhibit a reduced exciton--photon coupling strength of $21.8$~meV.
At a high pump fluence value of $\sim 30~\upmu \mathrm{J/cm^2}$ (right panel), the coupling strength decreases to $18.3$~meV, and the polariton resonances significantly broaden making the LP/UP splitting barely distinguishable (see more data in Supplementary Note 5).

The experimentally observed pump fluence-dependent polariton response can be modeled using transfer matrix calculations with variable relaxation rates.
The results of such calculations, corresponding to the selected pump fluence values, are shown in Fig.~\ref{fig:Nonlinear}c.
In order to match the experimental data, in the calculations we use different values of exciton radiative decay rate $\gamma_\mathrm{r}$ decreasing from $1.8$~meV (left panel) to $1.3$~meV (middle panel) and further to $0.9$~meV (right panel).
These values roughly correspond to the square-root scaling of coupling strength with exciton radiative decay rate $g \propto \sqrt{\gamma_\mathrm{r}}$.
Additionally, in order to account for power-induced spectral broadening, we increase the non-radiative decay rate for excitons $\gamma_\mathrm{nr}$ from $10$~meV (left panel) to $15$~meV (middle panel) and further to $25$~meV (right panel).

\begin{figure*}[t]
    \centering
    \includegraphics[width=\textwidth]{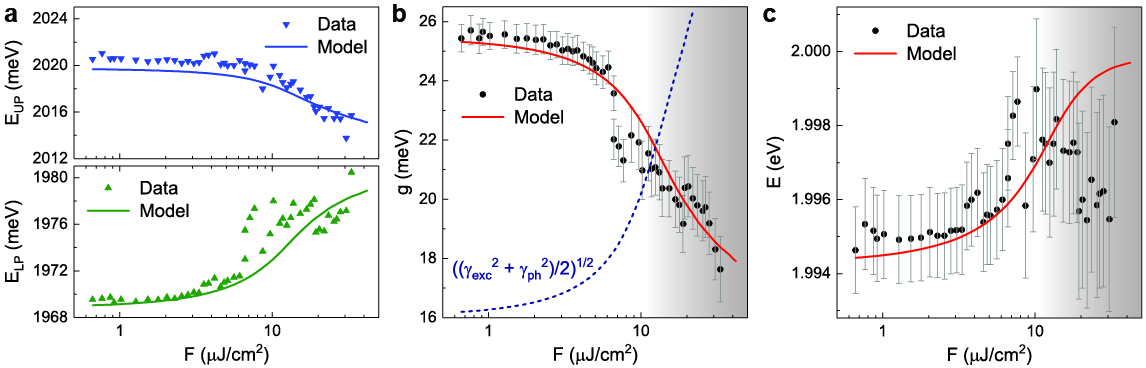}
    \caption{
    \textbf{Power-dependent parameters of guided exciton-polaritons in hBN/WS$_2$/hBN.}
    (a) Extracted from measurements (symbols) upper (top panel) and lower (bottom panel) polariton energies as functions of pump fluence, together with corresponding modeling results (curves).
    (b) Extracted from measurements light--matter coupling strength as a function of pump fluence (dots), together with corresponding modeling results (red solid curve) and effective relaxation rate for evaluating the strong light--matter coupling regime condition (blue dashed curve).
    (c) Extracted from measurements exciton energy as a function of pump fluence (dots), together with corresponding modeling results (curve).
    }
    \label{fig:Fluence}
\end{figure*}

Fig.~\ref{fig:Fluence} summarizes the trends observed in our pump fluence dependent experimental data.
The extracted spectral peak positions for the UP and LP polariton dispersion branches are plotted in Fig.~\ref{fig:Fluence}a.
These spectral positions are evaluated on resonance, that is, at the wavevector corresponding to zero detuning between the excitonic and photonic modes where the excitonic and photonic fractions in the polariton are 50$\%$.
While the UP energy (top panel, blue symbols) decreases with pump fluence, the LP energy (bottom panel, green symbols) increases exhibiting a somewhat stronger overall spectral shift.
This behavior translates into distinct pump fluence dependencies of the two parameters that we extract from experimental data via coupled oscillators model fitting: the exciton--photon coupling strength $g$ and exciton energy $E$.
As shown in Fig.~\ref{fig:Fluence}b, the coupling strength decreases by $\sim 8$~meV at the maximum fluence used in the experiment, corresponding to pump-induced collapse of Rabi splitting.
At the same time, the exciton energy exhibits a slight increase with increasing pump fluence (Fig.~\ref{fig:Fluence}c), with the maximum observed blueshift of $\sim 4$~meV.
Additionally, as presented in Supplementary Note 6, the exciton relaxation rate $\gamma_\mathrm{exc}$ increases up to $\sim 40$~meV for the experimental range of fluence values, resulting in pump-induced spectral broadening of polariton resonances.

We attribute the experimentally observed nonlinear optical response of polaritons in the hBN/WS$_2$/hBN sample to the underlying exciton collective effects within the WS$_2$ monolayer.
To theoretically describe the corresponding nonlinear energy shifts, we model the ultrafast dynamics of polariton population under pulsed near-resonant excitation, as explained in detail in Supplementary Note 7.
In our model, we take into account two effects.
The first effect is the reduction of exciton--photon coupling strength $g$ caused by excitation-induced filling of the available phase space and, as a result, the saturation of the interband exciton transition~\cite{brichkin2011effect}.
The second effect is the instantaneous renormalization of polariton energy caused by the exciton--exciton Coulomb interactions~\cite{ciuti1998role} and the associated spectral shift of the exciton energy $E$.

We calculate the corresponding theoretical values for the saturation rate of exciton--photon coupling $s = 21.45$ nm$^2$ and exciton--exciton exchange interaction rate $V_{\rm XX}=1.38 \, \upmu$eV$\cdot\, \upmu$m$^2$, which are close to previously reported theoretical and experimental estimates~\cite{Shahnazaryan2017,Barachati2018,Kravtsov2020,shahnazaryan2020tunable,bleu2020polariton,fey2020theory,Stepanov2021,erkensten2021exciton,cam2022symmetry,Benimetskiy2023,khestanova2024electrostatic}.
The results of modeling the pump fluence dependent parameters are shown in Fig.~\ref{fig:Fluence} with solid curves.
For the coupling strength dependence (b), the calculated value of $s$ provides good agreement between theory and experiment, with the corresponding saturation constant estimated as $V_\mathrm{sat} = g\cdot s \sim 0.55  \, \upmu$eV$\cdot\, \upmu$m$^2$.
This value is of the same order of magnitude but somewhat higher than what was reported recently for monolayer WS$_2$ in a Bragg mirror cavity at room conditions~\cite{Zhao2023}, which could be related to the excitation conditions being closer to the resonant energy and wavevector in our experiment. 
At the same time, to describe the experimental exciton energy dependence (c), we have to use a lower than the theoretical estimate effective exciton--exciton exchange interaction rate of $V_{\rm XX}^{\rm eff}=0.41 \, \upmu$eV$\cdot\, \upmu$m$^2$.
This discrepancy could be potentially related to the effect of sample heating with femtosecond laser pulses, which would contribute some red shift and reduce the overall blue shift of the observed exciton resonance as compared to the theoretical prediction.

The pump-induced collapse of Rabi splitting, presented in Fig.~\ref{fig:Fluence}b, eventually results in a transition between the strong-coupling and weak-coupling regimes~\cite{Shapochkin2020}.
Based on the condition $g > \sqrt{(\gamma_\mathrm{ph}^2 + \gamma_\mathrm{exc}^2)/2} \simeq 16.1$~meV evaluated from linear reflection measurements and the observed rate of Rabi splitting collapse, the transition from strong to weak exciton--photon coupling would be expected to occur at pump fluence $F \gtrsim 50~\upmu \mathrm{J/cm^2}$.
However, as the exciton relaxation rate is found to substantially increase with pump fluence (see Supplementary Note 6), the right-hand side of the above inequality increases as well.
We attribute this increase to the exciton--exciton annihilation induced broadening~\cite{ciuti1998role,gribakin2021exciton} and approximate it as a linear function of pump-dependent exciton density.
The corresponding model curve found as the best fit of experimental data is shown in Fig.~\ref{fig:Fluence}b with the dashed blue line.
As a result, the strong to weak coupling transition happens at substantially lower fluence values of $F \gtrsim 10~\upmu \mathrm{J/cm^2}$ where the strong coupling condition starts to break down, as indicated with gray-color gradient in Fig.~\ref{fig:Fluence}b, c.

The demonstrated efficient nonlinear response of guided polaritons in an hBN/WS$_2$/hBN stack, which is dominated by pump-induced collapse of Rabi splitting and exciton resonance broadening, can provide an enabling mechanism for developing active all-van-der-Waals integrated photonic elements~\cite{lee2024ultra, Ren2025} such as modulators and switches.
In these elements, the propagation of guided polaritonic modes can be effectively controlled via laser pulses switching between strong and weak coupling regimes, with corresponding change of dispersion characteristics and spectral response on ultrafast time scales between that of polaritonic and photonic modes.
We note that the use of monolayer instead of multilayer TMDs~\cite{Lee2024} provides crucial advantages in terms of low optical loss, efficient direct-bandgap emission, and possibilities of electrostatic control via gating.
Combined with the low cost of fabricating van der Waals stacks and recent developments in hBN nanostructuring, the controllable and strongly nonlinear response of exciton-polaritons in hBN/WS$_2$/hBN structures makes them an attractive material platform for optical neural networks and photon-based quantum computers.

Additionally, the novel approach for nonlinear near-field spectroscopy of guided polaritons demonstrated in this work provides further opportunities for studying optical response and light--matter interactions in a wide range of chip-compatible wave-guiding structures and materials.
In comparison to scanning near-field optical microscopy~\cite{Hu2017, Mrejen2019, Moore2025}, it offers an extremely fast (ms-scale) detection of full polaritonic dispersion with high spectral and angular resolution, albeit in a limited range of wavevectors, which is still sufficient for probing exciton-polaritons in semiconducting van der Waals materials.
In comparison to approaches based on oil-immersion objectives~\cite{Shin2022, Canales2023}, it allows tuning the near-to-far-field coupling and optimization of light--matter interaction, together with a possible extension to low-temperature measurements.
The use of femtosecond laser pulses enables further straightforward implementation of time-resolved near-field spectroscopy of guided polaritons based on correlated photon counting and various pump--probe techniques.

In conclusion, in this work we have studied the nonlinear response of guided exciton-polaritons in a van der Waals waveguide consisting of an hBN-encapsulated WS$_2$ monolayer.
To probe the nonlinearity associated with waveguide modes below the light line, we applied a novel near-field spectroscopic technique based on the combined use of a GaP SIL with controllable SIL--sample gap and femtosecond laser excitation.
We have experimentally demonstrated the formation of guided exciton-polariton modes with large Rabi splitting of $\sim 50$~meV and optical control of the associated exciton--photon coupling strength and polariton spectra, with a pump-induced transition from strong to weak coupling. 
The results of our study, supported with a theoretical model of polariton interactions in hBN/WS$_2$/hBN samples, suggest that the efficient nonlinear response of guided exciton-polaritons provides new opportunities for developing active integrated photonic devices based on van der Waals materials. \\


\begin{acknowledgments}
Sample fabrication was supported by Priority 2030 Federal Academic Leadership Program.
Optical measurements were supported by Russian Science Foundation project 25-72-20030.\\
\end{acknowledgments}

The data that support the findings of this study are available from the corresponding author upon reasonable request.


%

\end{document}